\documentclass{aa}

\usepackage{natbib}
\bibpunct{(}{)}{;}{a}{}{,} 
\input{bib.def}
\usepackage{graphicx}
\usepackage[colorlinks=true,urlcolor=blue,citecolor=blue,linkcolor=blue]{hyperref}
\usepackage{txfonts}
\usepackage{xcolor,colortbl}
\usepackage{stfloats}

\begin{document} 

   \title{Sausage, kink, and fluting MHD wave modes identified in solar magnetic pores by Solar Orbiter/PHI} 

      \author{S. Jafarzadeh \inst{1}
          \and
          L. A. C. A. Schiavo \inst{2,3} 
          \and
          V. Fedun \inst{4}
          \and
          S. K. Solanki \inst{1}
          \and
          M. Stangalini \inst{5} 
          \and
          D. Calchetti \inst{1}
          \and
          G. Verth \inst{6}
          \and
          D. B. Jess \inst{7,8}
          \and \\
          S. D. T. Grant \inst{7}
          \and
          I. Ballai \inst{6}
          \and
          R. Gafeira \inst{9}
          \and
          P. H. Keys \inst{7}
          \and
          B. Fleck \inst{10}
          \and
          R.~J.~Morton \inst{3}
          \and
          P. K. Browning \inst{2}
          \and
          S. S. A. Silva \inst{4}
          \and
          T.~Appourchaux \inst{11} 
          \and
          A.~Gandorfer \inst{1} 
          \and
          L.~Gizon \inst{1,14} 
          \and
          J.~Hirzberger \inst{1} 
          \and
          F. Kahil \inst{1} 
          \and
          D.~Orozco~Su\' arez \inst{12} 
          \and
          J.~Schou \inst{1} 
          \and
          H.~Strecker \inst{12} 
          \and
          J.C.~del~Toro~Iniesta \inst{12} 
          \and
          G.~Valori \inst{1}
          \and
          R.~Volkmer \inst{13}
          \and
          J.~Woch \inst{1} 
          }
   
  \authorrunning{Jafarzadeh {et~al.}}
   \institute{Max Planck Institute for Solar System Research, Justus-von-Liebig-Weg 3, 37077 G\"{o}ttingen, Germany\\
   \email{shahin.jafarzadeh@mps.mpg.de}
            \and
            School of Physics \& Astronomy, The University of Manchester, Oxford Road, M13 9PL, Manchester, UK
            \and
            Department of Mathematics, Physics and Electrical Engineering, Northumbria University, Newcastle upon Tyne, NE1 8ST, UK
            \and
            Plasma Dynamics Group, Department of Automatic Control and Systems Engineering, The University of Sheffield, Mappin Street, Sheffield, S1 3JD, UK
            \and
            ASI Italian Space Agency, Via del Politecnico snc, I-00133 Rome, Italy
            \and
            Plasma Dynamics Group, School of Mathematics and Statistics, The University of Sheffield, Hicks Building, Hounsfield Road, Sheffield, S3 7RH, UK
            \and
            Astrophysics Research Centre, School of Mathematics and Physics, Queen’s University Belfast, Belfast, BT7 1NN, UK
            \and
            Department of Physics and Astronomy, California State University Northridge, Northridge, CA 91330, USA
            \and
            Instituto de Astrof\'isica e Ci\^{e}ncias do Espa\c{c}o, Department of Physics, University of Coimbra, 3040-004, Coimbra, Portugal
            \and
            ESA Science and Operations Department, c/o NASA Goddard Space Flight Center, Greenbelt, MD 20771, USA 
            \and
            Univ. Paris-Sud, Institut d’Astrophysique Spatiale, UMR 8617, CNRS, B\^ atiment 121, 91405 Orsay Cedex, France
            \and
            Instituto de Astrof\'isica de Andaluc\'a (IAA-CSIC), Apartado de Correos 3004, E-18080 Granada, Spain
            \and
            Leibniz-Institut f\"ur Sonnenphysik, Sch\" oneckstr. 6, D-79104 Freiburg, Germany
            \and
            Institut f\"ur Astrophysik und Geophysik, Georg-August-Universit\"at G\"ottingen, Friedrich-Hund-Platz 1, 37077 G\"ottingen, Germany
} 
 
\def\corrAuthor{SJ}

\date{  }

\abstract
{Solar pores are intense concentrations of magnetic flux that emerge through the Sun's photosphere. When compared to sunspots, they are much smaller in diameter and hence can be impacted and buffeted by neighbouring granular activity to generate significant magnetohydrodynamic (MHD) wave energy flux within their confines. 
However, observations of solar pores from ground-based telescope facilities may struggle to capture subtle motions synonymous with higher-order MHD wave signatures due to seeing effects produced in the Earth's atmosphere. Hence, we have exploited timely seeing-free and high-quality observations of four small magnetic pores from the High Resolution Telescope (HRT) of the Polarimetric and Helioseismic Imager (PHI) on board the Solar Orbiter spacecraft, during its first close perihelion passage in March 2022 (at a distance of 0.5~au from the Sun). Through acquisition of data under stable observing conditions, we have been able to measure the area fluctuations and horizontal displacements of the solar pores. Cross correlations between perturbations in intensity, area, line-of-sight velocity, and magnetic fields, coupled with the first-time application of novel Proper Orthogonal Decomposition (POD) techniques on the boundary oscillations, provide a comprehensive diagnosis of the embedded MHD waves as sausage and kink modes. Additionally, the previously elusive $m=2$ fluting mode is identified in the most magnetically isolated of the four pores. An important consideration lies in how the identified wave modes contribute towards the transfer of energy into the upper solar atmosphere. We find that the four pores examined have approximately 56\%, 72\%, 52\%, and 34\% of their total wave energy associated with the identified sausage modes, and around 23\%, 17\%, 39\%, and 49\% to their kink modes, respectively, while the first pore also has around an 11\% contribution linked to the fluting mode. This study marks the first-time identification of concurrent sausage, kink, and fluting MHD wave modes in solar magnetic pores.
}


\maketitle

\section{Introduction}
\label{sec:introduction}

Theoretical models have suggested the existence of a large number of magnetohydrodynamic (MHD) wave modes in the same magnetic structures in the solar atmosphere \citep{1983SoPh...88..179E, 2005LRSP....2....3N, 2019mwsa.book.....R}, each of which can contribute to the heating of its outer layers \citep{2015SSRv..190..103J, 2015LRSP...12....6K, 2020ApJ...892...49H, 2021RSPTA.37900172G}. As such, identification of different MHD wave modes are essential for better understanding the energy budget of the upper atmosphere \citep[see, e.g.][and references therein]{2023LRSP...20....1J}. MHD waves are often generated in the low photosphere, across a variety of magnetic-field concentrations (on different spatial scales), and subsequently propagated into the upper solar atmosphere along the field lines \citep[e.g.][]{2006ApJ...648L.151J, 2011A&A...534A..65S, 2013ApJ...768...17M, 2017ApJS..229...10J, 2022ApJ...930..129B}.

Multiple MHD wave modes have only recently been identified in relatively large solar magnetic structures, such as pores and sunspots, thanks to sophisticated analysis approaches, such as k-$\omega$ filtering \citep{1988ESASP.286..315T, 1989ApJ...336..475T, 2003A&A...407..735R, 2017ApJ...842...59J}, B-$\omega$ analysis \citep{2021A&A...649A.169S}, as well as solar applications of Proper Orthogonal Decomposition (POD) and Dynamic Mode Decomposition (DMD) techniques \citep{2021RSPTA.37900181A}, applied to state-of-the-art high-resolution observations, with the results interpreted, in tandem, with theoretical/numerical models \citep[see also][]{2022NatCo..13..479S,2022ApJ...927..201A,2023ApJ...954...30A}. These studies have often concentrated on oscillations in intensity, line-of-sight (LOS) velocity, and polarisation signals within the magnetic structures \citep{2018ApJ...869..110S, 2020NatAs...4..220J, 2021RSPTA.37900216S, 2022ApJ...938..143G}. Additionally, the area perturbations, as a measure of MHD sausage modes \citep{2009A&A...494..295E, 2013A&A...551A.137M, 2013A&A...555A..75M} have also been explored in a few studies of sunspots and pores in photospheric time-series of intensity images \citep[e.g.][]{2014A&A...563A..12D, 2015A&A...579A..73M, 2016ApJ...817...44F, 2020RAA....20..117F}. By employing empirical decomposition methods on time variations of the size of magnetic pores, \citet{2011ApJ...729L..18M} found multiple signatures of magnetoacoustic sausage modes with periods in the range of 30--450~s. Later, \citet{2015ApJ...806..132G} reported the detection of upwardly propagating (slow surface) sausage modes in a magnetic pore, with a period range of 181--412~s, where a direct indication of wave energy dissipation was also observed. Furthermore, \citet{2018ApJ...857...28K} identified both surface and body sausage modes in several magnetic pores in the photosphere, with frequencies in the range 2--12~mHz and mean energy fluxes on the order of 6--43~kW/m$^2$ and 8~kW/m$^2$, respectively.

Area perturbations associated with MHD wave modes can more readily be measured in small-scale magnetic structures, due to the larger induced fractional variations in area  (e.g. small magnetic pores versus relatively large pores and sunspots). In addition to pores and sunspots, identification of sausage modes in the lower solar atmosphere has been reported through a number of studies in, for instance, small (point-like) magnetic elements \citep{2001A&A...371.1137B, 2014ApJ...796...72J, 2021RSPTA.37900175N, 2021SoPh..296..184G, 2023A&A...671A..69G}, fibrillar structures \citep{2012NatCo...3.1315M, 2017ApJS..229....7G, 2018ApJ...853...61S}, as well as in coronal loops \citep[e.g.][]{2003A&A...412L...7N, 2004ApJ...600..458A, 2020SSRv..216..136L}. 

In addition to sausage modes, transverse kink modes have often been simultaneously detected in a number of the above-mentioned studies \citep[see also][]{2015A&A...577A..17S,2017ApJS..229....9J}. In fact, work has shown that both sausage and kink waves may readily coexist in the same magnetic concentration and can interact with each other due to the coupling of their oscillatory motions \citep{2012ApJ...744L...5J, 2015A&A...579A.127L}. 

Sausage modes are characterised by the contraction and expansion of a flux tube along its axis, due to variations in the plasma density and magnetic field within the structure (e.g. a magnetic pore) as the wave passes along it \citep{1981SoPh...69...39R}. This results in periodic changes to the cross-sectional area of the magnetic concentration, which is also often associated with temperature fluctuations \citep{2009ApJ...702.1443F}. Such compressible waves are believed to have a large contribution to the heating of the upper solar atmosphere \citep{2012NatCo...3.1315M}. On the other hand, kink modes involve oscillations of the magnetic field lines, causing the magnetic structure to oscillate transversely \citep{1982SoPh...75....3S}. The (nearly) incompressible nature of the kink modes make it relatively difficult to dissipate their energy compared to, for instance, sausage modes \citep{1965RvPP....1..205B}. Various mechanisms, including buffeting by and/or interacting with external granules \citep{1990ApJ...348..346E, 1999ApJ...519..899H} and vortex-type photospheric drivers \citep{2011ApJ...727L..50K, 2018MNRAS.480.2839L}, have been shown to excite sausage and kink modes in magnetic flux tubes.

While mode conversion can occur close to the plasma-$\beta\approx1$ regions \citep[i.e. where the gas and magnetic pressures nearly coincide;][]{2003ApJ...599..626B, 2007AN....328..286C}, the coupling between sausage and kink modes may also take place when one mode induces or excites the other (\citealt{2000SoPh..193..139R}; cf. \citealt{2006A&A...446.1139V}). For example, the compression and expansion associated with the sausage modes may cause a change in the equilibrium magnetic field configuration within the flux tube, leading to a perturbation of the magnetic field lines. This perturbation can then excite the kink modes. Conversely, the kink modes can also influence the sausage modes. The transverse oscillations induced by the kink modes can cause changes in the plasma pressure and magnetic tension within the flux tube, resulting in modifications to the contraction and expansion behaviour of the sausage modes. 
Furthermore, interaction between the various MHD wave modes in a flux tube can be complex, potentially leading to the excitation of higher-order wave modes, such as fluting modes, or the modification of the existing modes. For example, an initially imposed kink mode in a thin flux-tube has been shown to excite a combination of sausage and fluting modes as part of the tube boundary perturbation \citep{2022SoPh..297..116R}.
Such couplings can affect the propagation characteristics, energy transfer, and damping rates of the oscillations within the flux tube. As such, the presence of these various MHD wave modes, particularly the fluting modes, may suggest a complex interplay between the magnetic fields, plasma dynamics, and the surrounding environment. Hence, identification of concurrent wave modes and their possible couplings are essential for better understanding of the dynamics and heating of the solar atmosphere. For a detailed mathematical description and further explanations of the various MHD wave modes, we refer to \citet{2019mwsa.book.....R}.

Besides depending on the size of the structure and spatial resolution of the observations, the identification of wave modes is affected by the variable seeing due to the Earth's atmosphere \citep{2021NatAs...5....5J}. As such, seeing-free observations of (spatially resolved) small-scale magnetic structures from space can better guarantee the absence of spurious signals and/or disturbances arising from the Earth atmospheric turbulence. 

In the present work, we analyse small magnetic pores in seeing-free, high-quality, and stable observations from the High Resolution Telescope \citep[HRT;][]{2018SPIE10698E..4NG} of the Polarimetric and Helioseismic Imager \citep[PHI;][]{2020A&A...642A..11S} on board the Solar Orbiter \citep[SO;][]{2020A&A...642A...1M} spacecraft. These small (yet spatially resolved) magnetic structures were found to present area and horizontal-displacement fluctuations, which are ideal cases for studying MHD (sausage and kink) wave modes. As such, oscillations in various physical parameters, namely area, intensity, LOS velocity and the longitudinal component of the magnetic field, are examined in four small (isolated) pores observed during the first perihelion of the nominal science phase of the SO in March 2022. 
Proper orthogonal decomposition (POD) techniques are employed to segregate the observed signals into constituent wave modes, where the associated plasma and oscillatory properties can be further characterised from the identified modes.

The observational data and their analyses are described in Sections~\ref{sec:observations} and \ref{sec:analysis}, respectively, with concluding remarks on the detected multiple wave modes being presented in Section~\ref{sec:conclusions}.

\section{Observations}
\label{sec:observations}

The data analysed in the present work were obtained by the SO/PHI-HRT during its first remote-sensing window of the nominal mission phase \citep{2020A&A...642A...3Z} on 2022~March~7 between 00:00--00:45 UTC, with a cadence of 60~s and a spatial sampling of $0{\,}.{\!\!}''5$ per pixel  (corresponding to 181~km on the solar surface at a distance of 0.501~au from the Sun). The observations captured the active region AR12960, consisting of several sunspots and pores of different sizes and properties, located at a cosine of heliocentric angle ($\mu$) of 0.87.
The HRT sampled an entire field-of-view (FoV) of $370\times370$ Mm$^2$ in full Stokes, employing the magnetically sensitive Fe~{\sc i}  6173.34~\AA\ line at six wavelength positions (five within the line and one in the continuum).

The full Stokes data, reduced on the ground \citep{2022SPIE12189E..1JS}, were corrected for optical aberrations introduced by the entrance window of the instrument without reconstructing for the diffraction at the entrance pupil \citep{2022SPIE12180E..3FK, 2023A&A...675A..61K}, 
resulting in a noise level of ${\approx1.8\times10^{-3}}$ in Stokes $Q/I_{c}$ and $V/I_{c}$, and ${\approx2.2\times10^{-3}}$ in Stokes $U/I_{c}$, where $I_{c}$ is the continuum intensity level. 
Furthermore, various physical parameters were inferred by means of Stokes inversions, using the Milne-Eddington C-MILOS code \citep{2007A&A...462.1137O}. 

Here, we analyse LOS velocities retrieved from the Stokes inversions, as well as Circular Polarisation (CP) as a measure of the longitudinal component of the magnetic field. Here, the CP parameter is calculated as, 
\begin{equation}
\label{eq:CP}
\mathrm{CP} = \frac{1}{4} \left( \sum_{i=1}^2 V_i - \sum_{i=4}^5 V_i \right) \ ,
\end{equation}
where $V_i$ are the Stokes $V$ parameters in the blue ($i=[1,2]$) and red ($i=[4,5]$) wings of the spectral line. The sign of the Stokes $V$ in the red wing was changed to avoid cancellation of opposite polarities in the two wings \citep{2011SoPh..268...57M}. The line-core position ($i=3$) was excluded due to the presence of mixed polarities. Thus, the CP has a reduced noise level (by about a factor of 2) compared to that from any single wavelength position in Stokes $V$.

For more information about these observations and their reduction/preparation procedures, as well some discussions on the preferred use of CP over the inferred magnetic field for wave studies, we refer the reader to \citet{2023A&A...674A.109C}, where the same dataset has been described in  greater detail.

\section{Analysis and results}
\label{sec:analysis}

We aim to investigate oscillatory signals, not only in intensity, LOS velocity and CP, but particularly in the size (area) of magnetic structures. Only small-scale magnetic flux tubes may present such (measurable) large fractional area variations within a relatively short period of time (i.e. the 45-min length of the time series of observations). We focus on the small magnetic pores in our observations, that are large enough to be spatially resolved, yet small enough to potentially show large fractional area variations. 

Within the entire FoV of the active region (see Figure~1 in \citealt{2023A&A...674A.109C}; also for the other data products) there are several pores of various sizes and properties. We have found only four candidates that, in addition to their small dimensions (with diameters on the order of 0.7~Mm), do not show any peculiar temporal evolution, such as interacting with neighbouring pores and/or other features over the course of the time series. 

\begin{figure*}[thp!]
    \centering
    \includegraphics[width=1.0\textwidth]{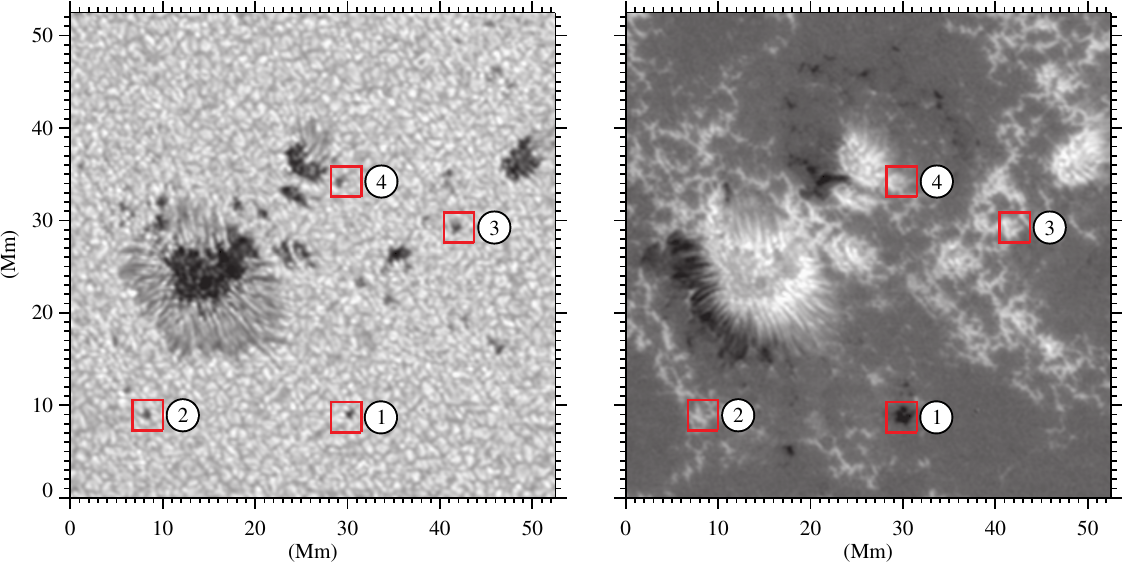}
    \caption{Four small pores of interest marked on a Stokes $I$ continuum image (left) and a CP map (right; with a range of [-2.9,5.7]\% in the units of Stokes $I$ continuum), corresponding to the middle of the time series.}
    \label{fig:FoV}
\end{figure*}

Figure~\ref{fig:FoV} illustrates a small part of SO/PHI-HRT's FoV in both Stokes~$I$ continuum (left) and CP (right), on which the four pores of interest are marked with small squares and numbered. The pores are present over the entire duration of the observations. However, pore number (2) displays a complex evolution towards the end of the time series, hence, it is analysed only over the first 37~min of the data set. Visual inspection of Figure~\ref{fig:FoV} indicates that the four pores are located in somewhat different magnetic environments, which can influence the local plasma flows \citep{2016RAA....16...78J}. 
It is evident that pore (1) is the most isolated one among our four features of interest, with no other pores, sunspots, or even many plage patches, nearby. Pores (2) and (3) are situated in the vicinity of other pores and both are located in the middle of plage/enhanced-network regions, where inflows from opposite directions may have produced the relatively high number density of magnetic-field concentrations in a small region. Finally, pore (4) is in a somewhat different situation compared to the other three, located in the immediate vicinity of a forming sunspot. From a close inspection of the time-series of images, it is also evident that pore (4) moves the largest distance (towards the forming sunspot) compared with the other three that have smaller spatial displacements. Furthermore, pore (1) has an opposite polarity compared to the other three features of interest.
It is worth noting that continuum intensity images are also essential for identifying such pores, in addition to magnetograms. For instance, pores (2)--(4) are located within plage/enhance network regions, hence they could hardly be detected as a pore from the CP map alone.

Each pore is analysed separately within individually selected sub-fields of the FoV marked with the red squares in Figure~\ref{fig:FoV}. The boundary of the pores at each time step is determined by means of an active contour segmentation method, applied on inverted mean-subtracted intensity images (of the Stokes~$I$ continuum; brightness values are inverted to facilitate segmentation). The most optimum contour levels were found to be at around 60\% of the maximum intensity (of the inverted mean-subtracted images). Slightly different contour/intensity levels have no effect on the oscillatory signatures, but provide small changes to the absolute values of the internal areas. After identifying the boundaries (i.e. the sizes of the pores), the continuum intensity, LOS velocity, and CP values are also extracted at all pixels within the pore boundaries. 
We note that LOS velocities across the four pores have a mean of 0.7 km/s and a standard deviation of 0.4 km/s. This is well below the spectral sampling of SO/PHI (70~m\AA, equivalent to $\approx3.4$~km/s), thus ensuring negligible influence on the CP measurements. Furthermore, we determine the centroid of each pore (the geometric centre of the pore) as the position of the magnetic structure at any given time, resulting in the calculation of (instantaneous) horizontal velocities of the pore over the length of the time series. The horizontal velocities were computed as transversal displacements of the pore within consecutive frames divided by the cadence of the observations. These have all together provided us with the possibility of exploring the temporal evolution (fluctuations) in the area, horizontal velocity, mean intensity, mean LOS velocity, and mean CP of each pore (each of the latter three parameters were obtained as an average over the entire area of the pore). Since we are interested in wave signatures and not the slow evolution of the pores, we subtract low frequencies ($<1$~mHz) from all signals by means of wavelet filtering. All signals are also detrended (linearly) and apodized (using a Tukey window) prior to any further analyses.

\begin{figure}[!tp]
    \centering
    \includegraphics[width=0.49\textwidth]{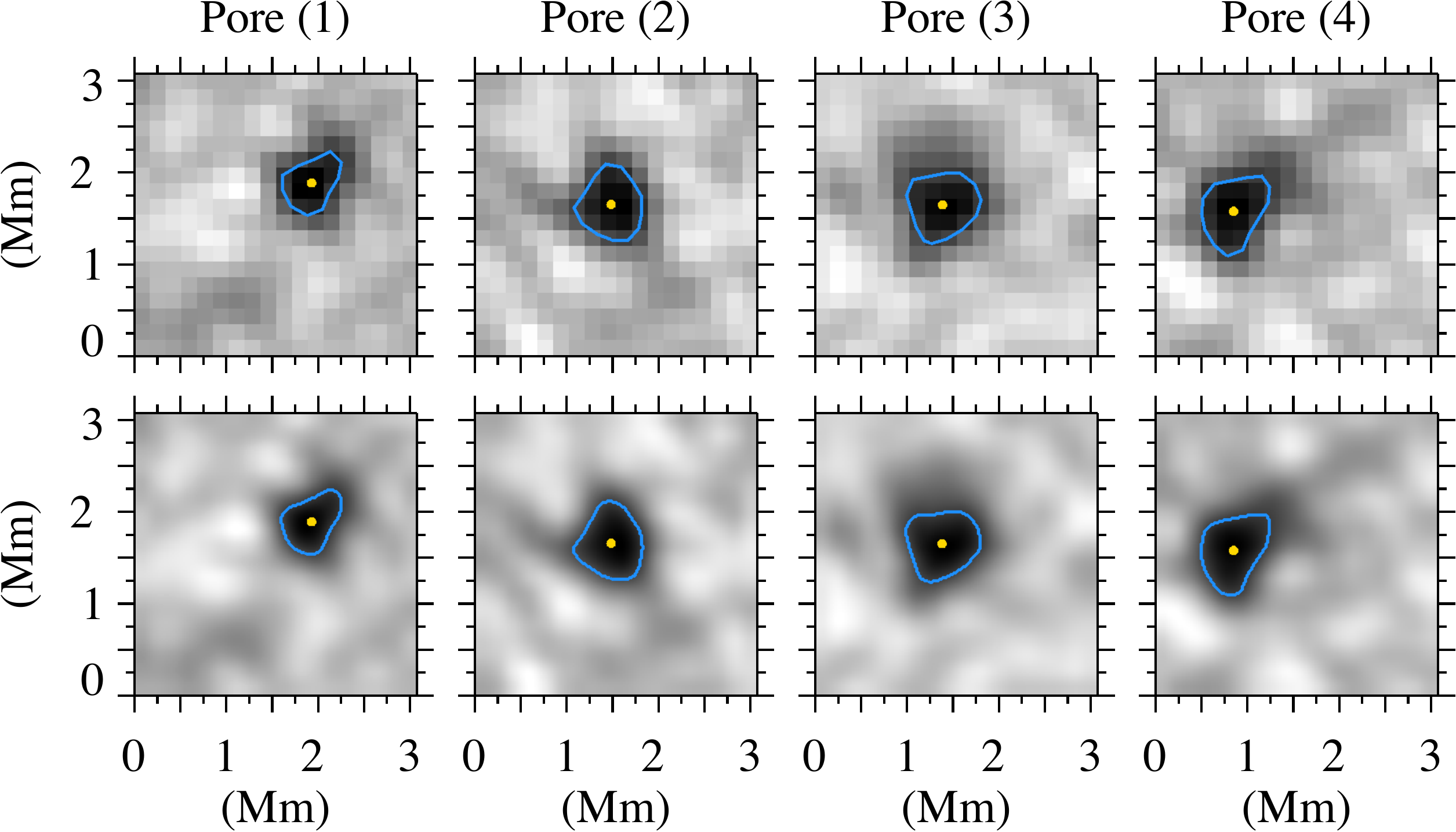}
    \caption{The four pores of interest (in the Stokes $I$ continuum) in both original spatial sampling (upper row) and a higher sampling resolution (lower row) in the middle of the time series (as marked in Figure~\ref{fig:FoV}). The blue solid-line contours depict the identified pore boundaries and the yellow dots mark the centroid of the pores.}
    \label{fig:pores}
\end{figure}

\begin{figure*}[hpt!]
    \centering
    \includegraphics[width=\textwidth]{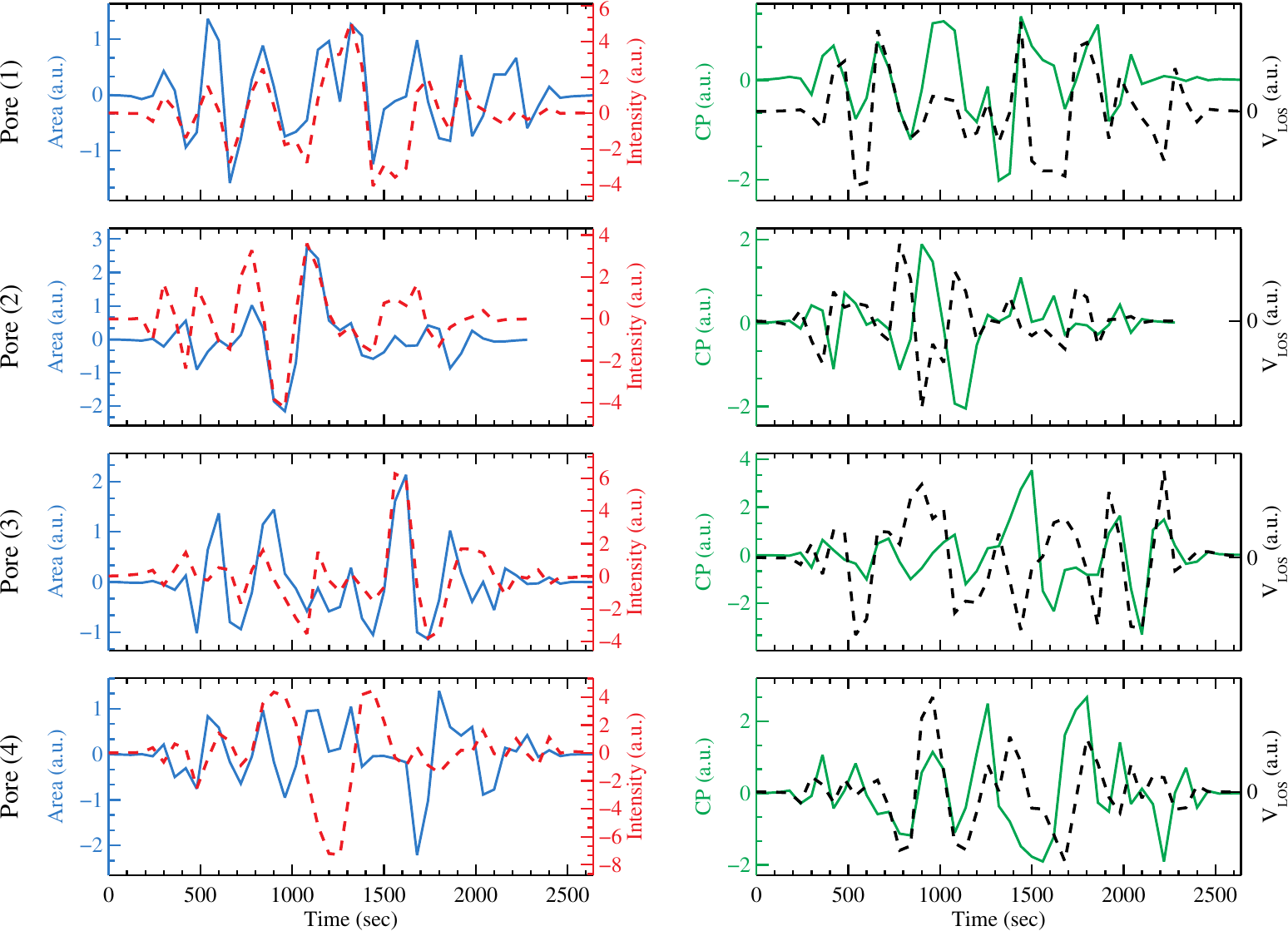}
    \caption{Perturbations in various physical parameters of the four pores of interest. The left panels show the area (blue solid) and pore-averaged intensity (red dashed) fluctuations, whereas the pore-averaged CP (green solid) and pore-averaged LOS velocity (black dashed) oscillations are presented on the right panels. The plots in a given row show the results for one pore, as indicated on the left of the figure. All signals are detrended (linearly) and apodized (using a Tukey window). Low frequencies (<1~mHz) have been filtered out. As such, the amplitudes are not preserved and thus they are all shown in arbitrary units (a.u.).}
    \label{fig:oscillations}
\end{figure*}

Figure~\ref{fig:pores} represents the four pores in both original spatial-sampling of the observations (upper row) and a higher sampling resolution (increased by a factor of 10; using linear interpolations) for better visibility (lower row). We note that the analyses are performed on the original images, with the exception of modelling the boundary oscillations using the POD technique. Although the area oscillations are identical from both sets of spatial samplings (hence has no effect on the modelling), the higher sampling-resolution images facilitate mode identification through POD (see Section~\ref{sec:pod}).

\subsection{Perturbations in physical parameters}
\label{sec:perturbations}

The temporal variations of the area, pore-averaged intensity, pore-averaged LOS velocity, and pore-averaged CP are displayed in Figure~\ref{fig:oscillations} for the four pores of interest. Due to the high-pass frequency filtering, the original amplitudes, even at high frequency, might be slightly affected. This has no impact on the results, since we are only interested in identifying oscillatory signals and their phase relationships. Hence, all plots are shown in arbitrary units. Moreover, we primarily aim to study area fluctuations in the small magnetic pores, that together with phase relationships between other oscillatory signals, can better characterise the sausage, and possibly kink, modes. 

\begin{figure}[!ht]
    \centering
    \includegraphics[width=0.49\textwidth]{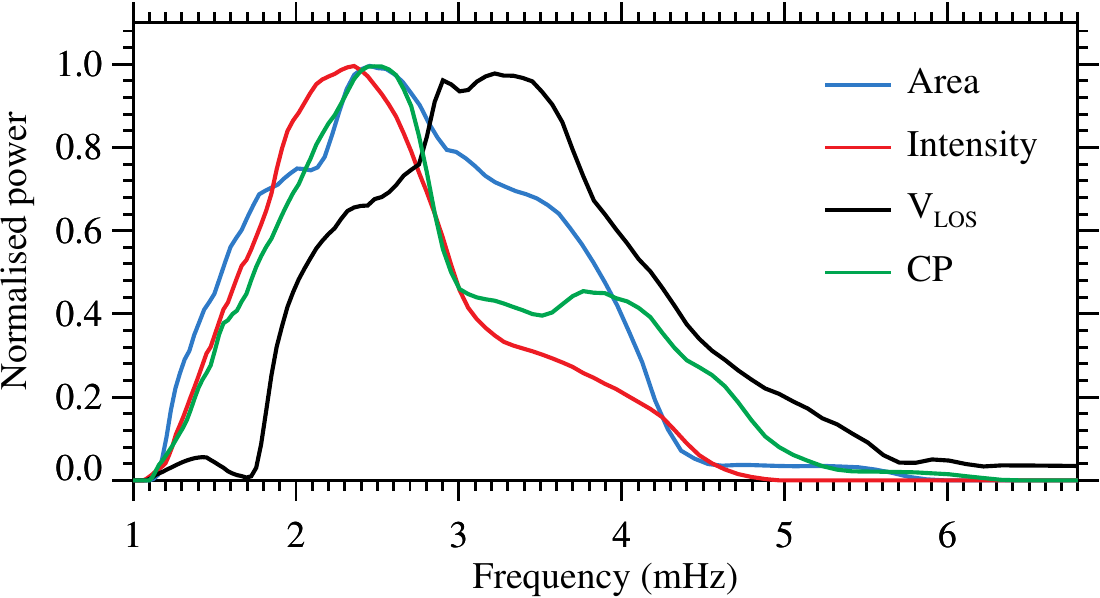}
    \caption{Modified global wavelet power spectra of the four observables shown in Figure~\ref{fig:oscillations} for the four small pores. The power spectra are normalised by their maximum value.}
    \label{fig:frequency_distributions}
\end{figure}

Figure~\ref{fig:frequency_distributions} presents modified global wavelet power spectra of the four observables shown in Figure~\ref{fig:oscillations}, combined for the four small pores (the four pores display similar power spectra for fluctuations in each of the observables). The modified global wavelet spectra are time-integrated power spectra that only include areas in the wavelet power spectrum (using a Morlet function; \citealt{1998BAMS...79...61T}) falling within the 95\% confidence levels and outside of the wavelet's cone-of-influence (CoI; subject to edge effects), in other words, they represent the wavelet's power-weighted frequency distributions with significant power, unaffected by the CoI. The oscillations in area, intensity, LOS velocity, and CP show a relatively wide frequency distributions with peaks at around 2.5, 2.3, 3.3, and 2.5~mHz, respectively. The small differences between the peaks could be representatives of different MHD wave modes, similar to those found by \citet{2023A&A...674A.109C} in other (considerably larger) magnetic structures within the same active region, from the same dataset (albeit using a different analysis approach).

\begin{figure*}[th!]\setlength{\hfuzz}{1.0\columnwidth}
\begin{minipage}{\textwidth}
     \centering
     \includegraphics[width=0.99\textwidth]{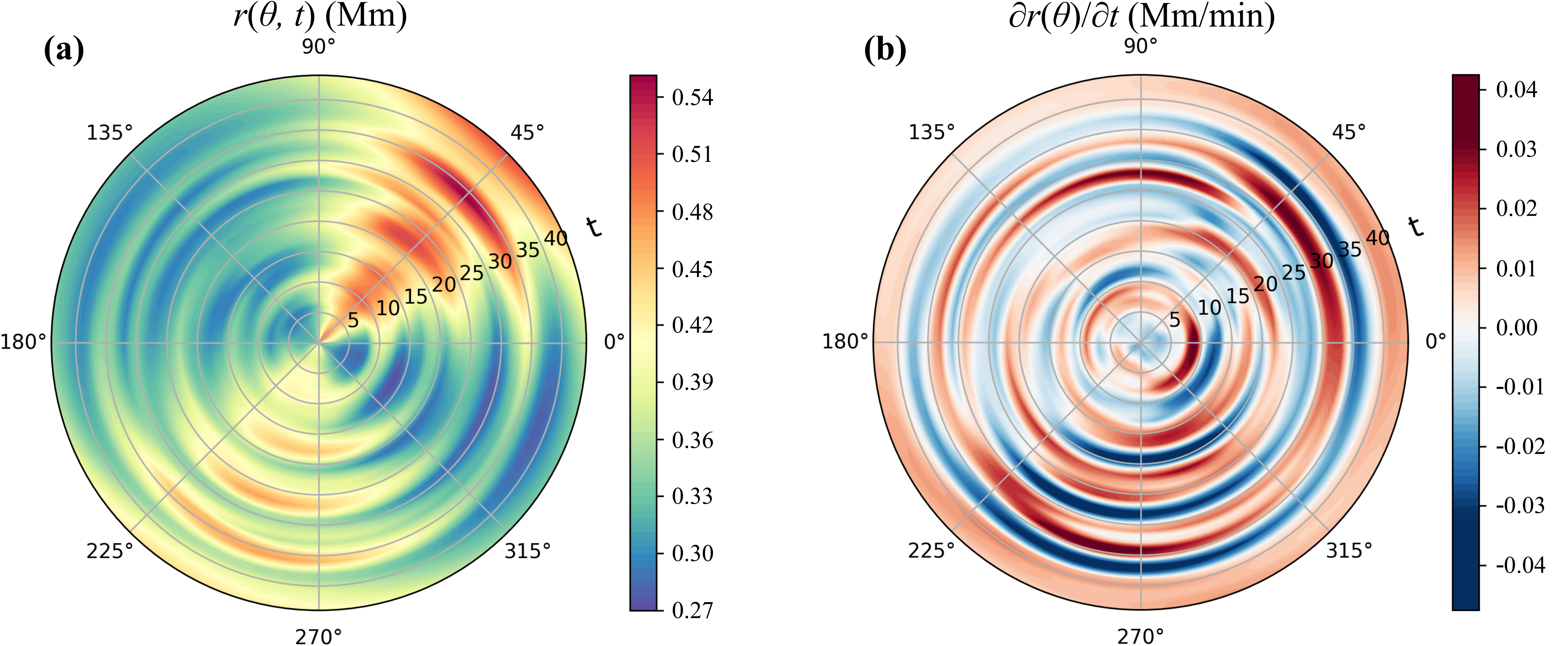}
     \caption{Contour maps of the pore (1)'s radial coordinates as a function of polar angle, $\theta$, and time, $t$ (a), and of $\partial r(\theta)/\partial t$ (b). The time variation is plotted in the radial axis (in minutes) for both contour maps.}
     \label{fig:modal_analysis}
\end{minipage}
\end{figure*}

The areas of the pores (1)--(4) have maximum variations (after detrending and high-pass filtering) of 33\%, 54\%, 39\%, and 31\%, respectively. The fact that we observe such large area oscillations most likely indicates the presence of sausage modes. Furthermore, we observe horizontal-velocity perturbations (with frequencies on the order of 2--5 mHz) which are not artefacts produced by the spacecraft jitter as the various pores are not moving in phase in (exactly) the same direction.
Albeit these horizontal-velocity fluctuations have relatively small amplitudes ($\approx0.7$~km/s on average), they may imply the presence of kink modes (such oscillations could be caused by the dynamics of the external flow, or they could be intrinsic to the pore).

To further examine these initial suggestions, we perform a modal analysis using the POD technique (Section~\ref{sec:pod}), followed by a wavelet phase-lag analysis between pairs of oscillations observed in the various parameters (i.e. those shown in Figure~\ref{fig:oscillations}), in comparison with theoretical models (Section~\ref{sec:[phase-analysis]}).

\subsection{Proper Orthogonal Decomposition (POD) of pore boundary shapes}
\label{sec:pod}

The large oscillations in pore areas, as shown in the previous Section \ref{sec:perturbations}, also naturally result in measurable changes in pore shape. Therefore, in this Section we aim to identify modes by analysing how the shape of the pore boundary changes with time.  We will decompose this motion using Proper Orthogonal Decomposition (POD).  POD is an sophisticated data-driven method which can extract dominant wave modes by identifying spatial patterns and temporal dynamics of a structure that significantly contribute to its overall variability. More specifically, POD determines the eigenfunctions that are orthogonal in space. As such, POD allows for the decomposition of a time series into spatial modes that are associated with a range of different frequencies or spectral bands.

POD is a powerful data-analysis method used to obtain a low-dimension approximation of a high-dimensional process. It was initially introduced in the context of fluid dynamics by \citet{Lumley1967} to analyse coherent structures in turbulent flows. Only recently has its application on large solar magnetic structures (i.e. sunspots) been developed, allowing for the first time the identification of multiple high-order eigenmodes in the photosphere \citep{2021RSPTA.37900181A, 2022ApJ...927..201A}.
POD is a particularly useful modal decomposition for inhomogeneous unsteady flows/environments. Its fundamental idea is to decompose a set of fluctuation fields, the so-called ``snapshots'', into a sum of orthonormal spatial modes organised/ranked by their eigenvalues (or their contribution to the total variance). The larger the eigenvalue, the more variance (or energy) is captured by that mode. This means that POD modes with larger eigenvalues are more significant in terms of explaining the variability in the data. Thus, both spatial structure and temporal evolution of each POD (or empirical) mode are obtained.

Before we analyse how the shape of the pore boundaries change with time to identify MHD wave modes, first we note that the oscillations of the pore centroids may be attributed to the movement of the external flows, which are highly non-stationary. For this reason, our analysis is carried out after shifting each pore's centroid to the centre of the FoV (i.e., the origin of the system) in an attempt to isolate the movements of the pore boundary from the advective effects of the external flow field.
Then, we extract the $x$ and $y$ coordinates of the boundary (for each magnetic pore) at each time step for the entire duration of the observations and determine the distance between the coordinates of the boundary and the position of the centre. This helps us to transform the shape of the pore into a polar coordinate system ($r,\theta$), interpolated into an equally spaced grid in the $\theta$ direction, with 360 points. 

This process is repeated for all instances in time, generating a matrix that describes the variation of the shape of the analysed pore as a function of time (i.e. $r(\theta,t)$) as displayed in Figure~\ref{fig:modal_analysis}a for pore (1). The plot illustrates the temporal variation of the size of the pore in all directions in polar coordinates, for the whole duration of the time series. Each circle represents the size of the pore (i.e. the distance between the pore boundary and the pore's centre) depicted with the background colour for all polar angles at a particular time step, with time starting from the centre of the circle going outwards, so the radius of the disk is the time of observations.
The resulting matrix is helpful for analysing the dynamical system and is used for the modal analysis of the pore. 

It can be seen in Figure~\ref{fig:modal_analysis}a that the background colour in the radial direction (representing the time variation of the pore's radius) is increasing and decreasing at all angles, showing that the pore is expanding and contracting (i.e. boundary oscillations). It is also evident that the fluctuation of the pore's radius is not uniform in the $\theta$ direction, predominating/peaking at around $\theta$ = 45$^\circ$ and 250$^\circ$, which may be related to a periodic movement of the centre of mass. We note that both the pore's radius and the oscillation's amplitude are larger at the abovementioned $\theta$ directions. In other words, the variation of the pore's radius is larger in the directions in which the pore is more extended.

Figure \ref{fig:modal_analysis}b shows $\partial r(\theta)/\partial t$. This plot reveals regular contractions and expansions taking place at all polar angles, with periods on the order of 7--10~min linked to this oscillatory process. 
These matrices, containing the coordinates of the pore boundary, are used as an input for the POD analysis.
In the present work, a snapshot refers to the shape of the boundary of each pore at one moment in time (Figure~\ref{fig:modal_analysis} represents a set of snapshots), and POD is able to compute as many empirical modes as there are time snapshots (i.e. 45). However, not all POD modes are physical. Many of them can be due to, for instance, noise or spurious signals. Hence, to identify which empirical modes describe MHD modes, the POD spatial modes are compared with those predicted by cylindrical flux-tube models (when amplitudes of the POD modes are considerably larger than the noise levels).
As such, we are decomposing the radial coordinate of a pore boundary as
\begin{equation}
r\left(\theta,t \right) \; = \; \left\langle r(\theta) \right\rangle 
+ r^{\prime}\left(\theta,t \right) \; = \; \left\langle r(\theta) \right\rangle 
+ \sum_{n=1}^{N} a^{\left( n \right)} \left(t \right) \phi^{\left( n \right)} \left(\theta \right)\mbox{ ,} 
\end{equation}
where $\phi^{\left( n \right)}$ represents a set of space-dependent orthonormal modes, $a^{(n)}$ is a time-dependent mode amplitude, $N$ is the number of snapshots and $n$ identifies the mode index. Here $\left\langle \, \, \right\rangle$ represents a time average and the {\it prime} denotes a fluctuation. 
A reconstructed fluctuation field can then be approximated by,
\begin{equation}
\tilde{r}^{\prime}\left(\theta,t \right) \; \approx \; \sum_{n=1}^{M} a^{\left( n \right)} \left(t \right) 
\phi^{\left( n \right)} (\theta) \mbox{ ,}
\end{equation}
where $M$ is the number of modes used in the reconstruction. Using the snapshots method introduced by \citet{Sirovich1987}, the modal basis is constructed using a covariance matrix of radius fluctuation field as,
\begin{equation}
C_{t_{1},t_{2}} \; = \; \frac{1}{N} \int_{\Omega} r^{\prime}\left(\theta,t_{1}  \right)  
r^{\prime} \left(\theta,t_{2}  \right) d\theta \mbox{ .}
\end{equation}

The above matrix is symmetric, positive, and semi-definite, so we can compute the eigenvalues and eigenvectors using singular value decomposition (SVD). Thus, the POD spatial modes can be computed by a linear combination of the snapshots into an orthonormal set of basis functions,
\begin{equation}
\phi^{\left( n \right)} \left(\theta \right) \; = \; 
\frac{1}{\lambda_{n}N} \sum_{k=1}^{N} \xi_{k,n} r^{\prime}  \left(\theta,t_{k}  \right) \mbox{ ,}
\end{equation}
where $\lambda_{n}$ are the eigenvalues and $\xi_{k,n}$ represent the sets of eigenvectors of the correlation matrix $C$. The term $k$ represents the $k$th column of $\xi$ in the eigenvalue problem $C\xi = \lambda\xi$. Finally, the time-dependent mode amplitude is given by, 
\begin{equation}
a^{\left( n \right)} \left(t_{k} \right) \; = \;  \sqrt{N \lambda_{n}} \xi_{k,n} \mbox{ .}
\end{equation}

\begin{figure}[!b]
     \centering
         \includegraphics[width=0.49\textwidth]{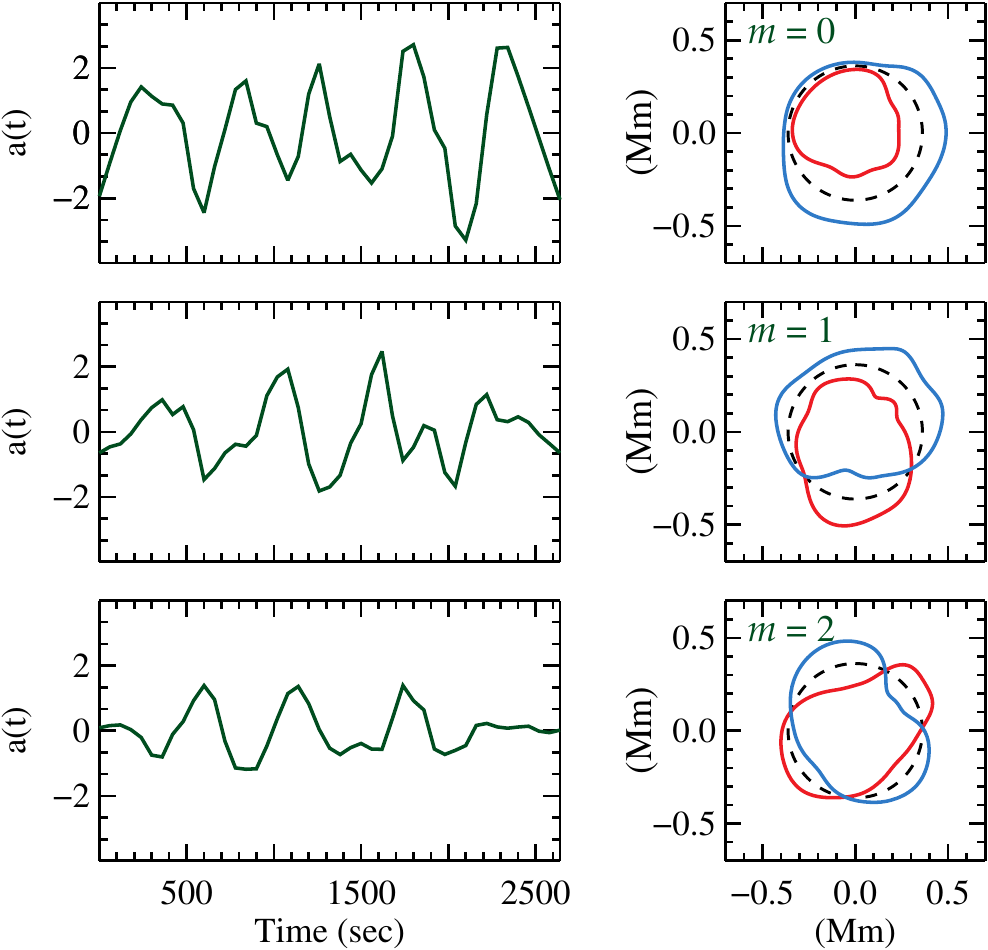}
         \caption{The left column illustrates the results of POD temporal modes, $a(t)$, for the first three modes  (i.e. sausage, kink, and fluting modes, corresponding to the azimuthal wave numbers, $m$ = 0, 1, and 2) for pore (1). The right column presents their corresponding spatial modes, $\phi(x,y)$, in a Cartesian coordinate system, projected over a reference circle. The blue and red curves show the maximum positive and negative values of the perturbations, respectively.
}
         \label{fig:pod_modes_pore1}
\end{figure}

Figure \ref{fig:pod_modes_pore1} illustrates the results of the modal analysis obtained by POD for pore (1). The left panels display the temporal evolution of the coefficients $a(t)$, and the right panels show their spatial structures for the first, second, and third POD modes (i.e. those with $n$ = 1, 2, and  3). Here, only the first three POD modes were found reliable and physical as the other modes do not present an oscillatory time coefficient or their contribution to the overall signal is insignificant.
The spatial perturbations, projected over a reference circle (depicted by a black dashed line), are retrieved from the following equations,
\begin{eqnarray}
    \label{eq:modal-shape}
    x &=& (1 \pm r^{\prime}) \ \cos(\theta) = (1 \pm \phi) \ \cos(\theta) \ , \\ \nonumber
    y &=& (1 \pm r^{\prime})\ \sin(\theta) = (1 \pm \phi) \ \sin(\theta) \ .
\end{eqnarray}
The spatial modes are shown in blue for positive and in red for negative values of disturbances. The temporal coefficients of the first three empirical modes for pore (1) reveal harmonic perturbations with peak frequencies at 2~mHz, 1.7~mHz, and 1.8~mHz, respectively. The first mode is predominantly due to the (asymmetric) radial expansion and contraction of the pore in all directions, thus representing a sausage mode (the azimuthal wave number, $m=0$) when compared with a cylindrical flux-tube model. The second mode is mainly related to the movement of the centre of mass along the $y$-axis, corresponding to a kink mode ($m=1$). Finally, the third POD mode is primarily representative of a fluting wave mode ($m=2$), according to the standard cylindrical flux-tube models. 
We note that the MHD modes are recognised by their primary mode characteristic, which refers to their main type of motion in our observations. It is important to note that the oscillatory movements seen in the pore cross-sections may not perfectly align with the expected movements for the eigenmodes of a cylindrical flux tube (that exhibit perfect symmetry, both in terms of shape and oscillations). This discrepancy can be attributed not only to the irregular, non-circular shapes of the observed waveguides, namely the small pores, but also to the presence of asymmetric oscillations in the observations. This variation, however, does not significantly impact the overall agreement, or correlation, between the observed and theoretical wave modes. In fact, the identification of each MHD wave mode is determined by the best agreement obtained, indicating the most accurate match between them.

\begin{figure}[!th]
     \centering
         \includegraphics[width=0.49\textwidth]{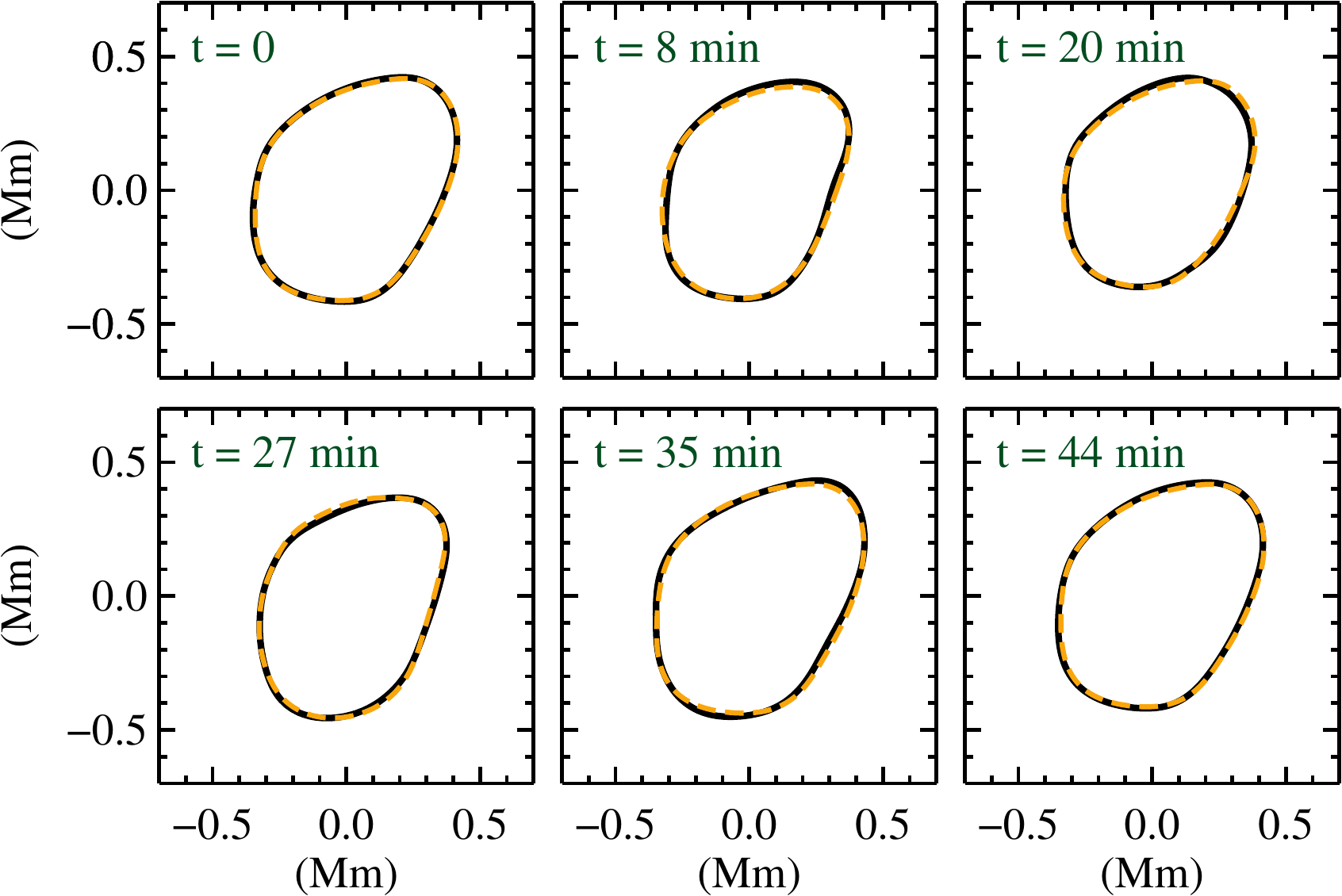}
         \caption{Comparison between the boundary of pore (1) at different time steps. The black solid lines represent the boundary extracted from the observational data, and the orange dashed lines depict POD reconstructions using the first three POD modes.}
         \label{fig:pore_reconstruction}
\end{figure}

Figure~\ref{fig:pore_reconstruction} illustrates how the reconstructions of a pore's boundary  (i.e. its shape) at six selected time steps from the first three modes detected by POD (orange dashed curves) match the identified boundaries from the observations (black solid curves). 
It is evident that all the POD reconstructions agree well with the original data and almost perfectly capture the pore dynamics (for the first three empirical modes).  
It worth noting again that POD is a data-driven technique, hence, it is not limited to predefined eigenfunctions, thus, POD has been able to reliably decompose MHD wave modes in the small pores with irregular cross-sectional shapes.

\begin{figure}[h]
     \centering
         \includegraphics[ width=0.49\textwidth]{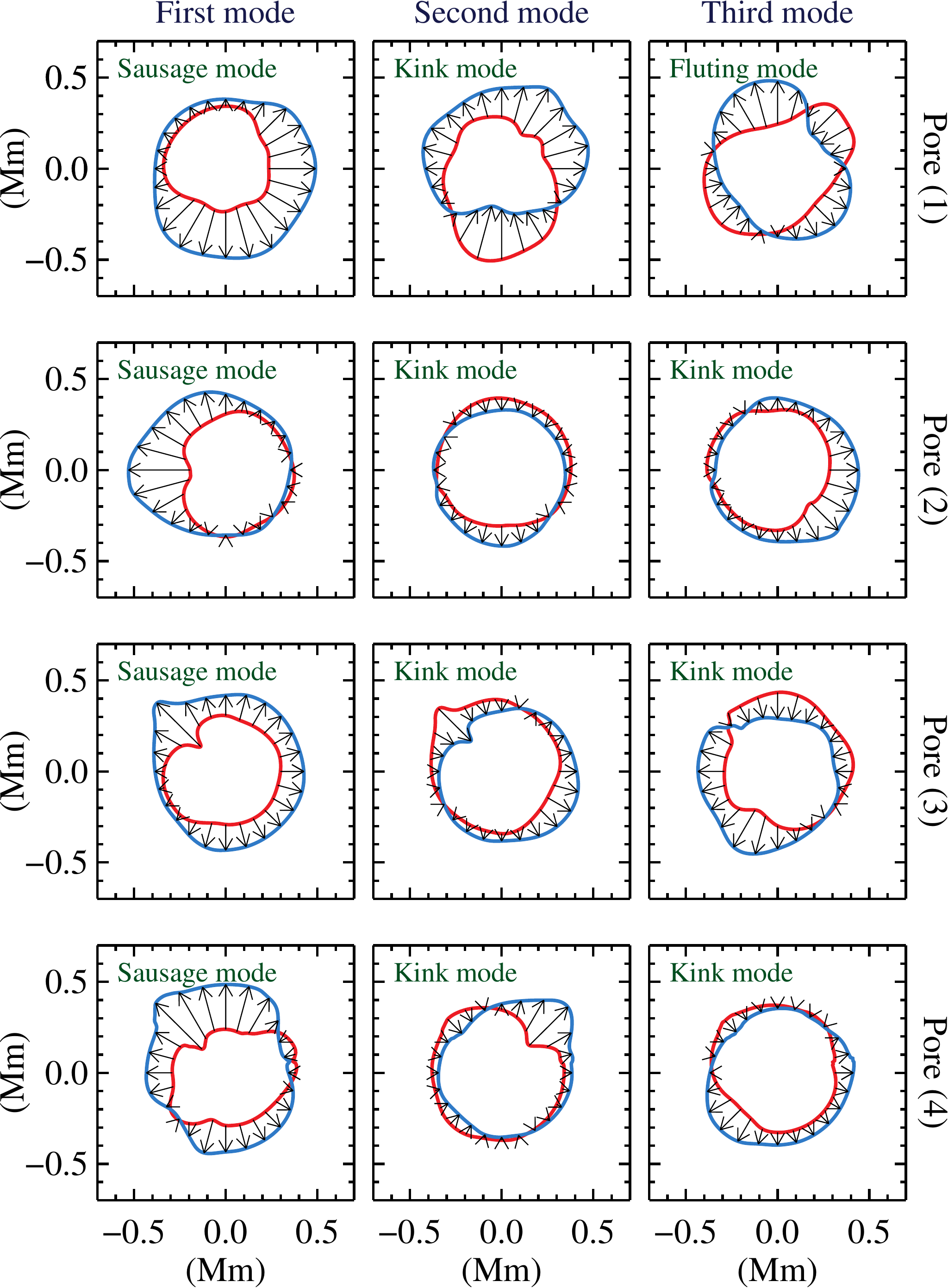}
         \caption{Modal perturbations of the four pores of interest recovered by POD. From left to right, the three columns illustrate the first, second, and third POD modes (ranked by their eigenvalue magnitudes) of each pore in a separate row. The blue and red lines depict the maximum positive and negative values of  perturbations, respectively. The arrows connect selected points from the red to blue curves to illustrate directions of motion within each half oscillation period. The identified eigenmodes are marked on the top-left corner of all panels. For all four pores, the first and second modes represent the sausage and kink modes, respectively. The third mode revealed a fluting mode for pore (1), and kink modes for pores (2)-(4). The temporal evolution of the modes for each pore are available as \href{https://sync.academiccloud.de/index.php/s/j55vX7mz4rkflS4}{online movies}.}
         \label{fig:pod_modes_all_pores}
\end{figure}

\begin{figure}[h]
     \centering
         \includegraphics[ width=0.48\textwidth]{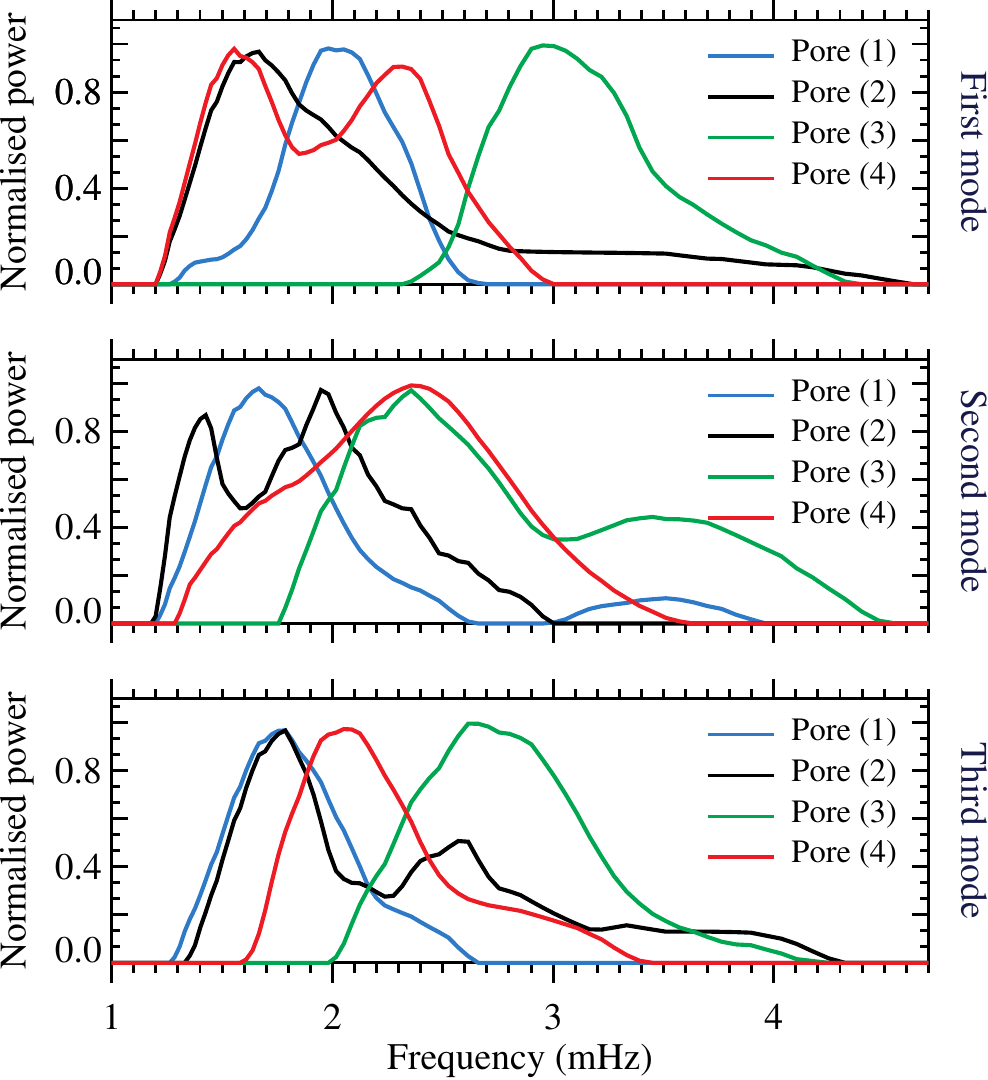}
         \caption{Modified global wavelet power spectra of the first three modes obtained by POD, for the four pores of interest.}
         \label{fig:pod_modes_frequencies}
\end{figure}

Figure~\ref{fig:pod_modes_all_pores} summarises the spatial structure of perturbations of the boundary of all four pores, as a result of the POD analysis applied to the remaining three pores (for completeness, it also includes again those of pore (1)). The three POD modes are organised in three columns from left to right, while rows denote one of the four pores as indicated on the right side of the figure. As in Figure~\ref{fig:pod_modes_pore1}, the blue and red curves depict the positive and negative values of perturbations, respectively. It is evident that for all the analysed pores, the sausage wave mode is the dominant POD mode followed by the kink mode. Thus, the nature of the first two energetic modes (i.e. those with the largest eigenvalues) is the same as in the case of pore (1), however unlike pore (1), which showed the existence of  a fluting mode, the third POD mode identified in pores (2)--(4) is predominantly a kink mode. This kink mode of lower contribution (i.e. with a smaller variance; associated to the third POD mode) might be due to the non-circular cross-section of the structure. Furthermore, the variance observed between the two kink modes, in each pore, could potentially be attributed to the difference in excitation strength. For example, granular buffeting might be stronger in one direction compared to its perpendicular, leading to the excitation of kink modes with perpendicular polarisations.
It is worth noting again that the MHD modes are identified by their main characteristics. However, a `flavour' of another wave mode may also be observed in some of the cases (as a result of, e.g. mode mixing), though such distinctions are not straightforward form the analysis of the boundary oscillations alone.

To better visualise the main dynamics of the oscillations, arrows connecting (equally distanced) selected points from the negative to positive fluctuations (i.e. from the red to blue curves) are also depicted in Figure~\ref{fig:pod_modes_all_pores}. 
These arrows indicate that the sausage modes (i.e. the first POD mode) are associated with the (asymmetric) expansion of the waveguides as the predominant motion in all four pores, while some small (minor) distortions are also observed in the case of pores (2) and (4). Moreover, for all four magnetic structures, percentage difference in area between the maximum negative and positive perturbations (associated to the first mode) are larger than 50\%.
Table~\ref{table:size_percentage_difference} lists the area percentage differences between the maximum negative and positive perturbations (i.e. between the blue and red curves in Figure~\ref{fig:pod_modes_all_pores}). 
In the case of kink modes (from both second and third POD modes), the main dynamics are characterised by (asymmetric) movement from one side to the other, while some partial expansions may also be observed in parts of some of the irregularly-shaped waveguides (that are not dominating). Additionally, their size variations are considerably lower compared to those in the sausage modes.
Finally, the fluting mode (the third POD mode of pore (1)) mainly displays oppositely directed motions along perpendicular lines, with relatively small size changes.

\begin{table}[hb]
\caption{Area percentage differences between the maximum negative and positive perturbations associated to the three wave modes in the four pores under study.} 
\label{table:size_percentage_difference}
\centering
\begin{tabular}{l c c c}
\hline\hline \\  [-1.9ex] 
 & First mode & Second mode & Third mode \\
\\ [-1.9ex]
\hline \\ [-2.0ex]              
    Pore (1) & 84\% & 28\% & 14\% \\
    Pore (2) & 50\% & 2\% & 36\% \\
    Pore (3) & 64\% & 7\% & 11\% \\
    Pore (4) & 67\% & 9\% & 22\% \\
\\ [-2.0ex]
\hline
\end{tabular}
\end{table}

The modified global wavelet power spectra of the three POD modes, for the four pores, are shown in Figure~\ref{fig:pod_modes_frequencies}, with their dominant frequencies summarised in Table~\ref{table:POD_frequencies}.

\begin{table}[ht]
\caption{Oscillations frequency of the POD modes identified in the four pores of interest.} 
\label{table:POD_frequencies}
\centering
\begin{tabular}{l c c c}
\hline\hline \\  [-1.9ex] 
 & First mode & Second mode & Third mode \\
\\ [-1.9ex]
\hline \\ [-2.0ex]              
   Pore (1) & 2.0~mHz & 1.7~mHz & 1.8~mHz \\   
   Pore (2) & 1.7~mHz & 1.4 and 1.9~mHz & 1.8~mHz \\
   Pore (3) & 3.0~mHz & 2.3~mHz & 2.7~mHz \\
   Pore (4) & 1.5 and 2.3~mHz & 2.4~mHz & 2.0~mHz \\
\\ [-2.0ex]
\hline
\end{tabular}
\end{table}

Thus, the pores show slightly different dominant frequencies of the different modes, within the ranges of 1.5--3.0~mHz for the sausage modes, and 1.4--2.7~mHz for the kink modes (from the second and third POD modes), and a frequency of 1.8~mHz for the single observed fluting mode. These differences in the frequencies of the same modes in the four pores can be attributed to several factors such as the transversal size of the magnetic waveguide, the wavelength of waves and the physical parameters that describe the state of the plasma (strength of the magnetic field, temperature, density, etc.).

\begin{figure}[h]
     \centering
         \includegraphics[width=0.49\textwidth]{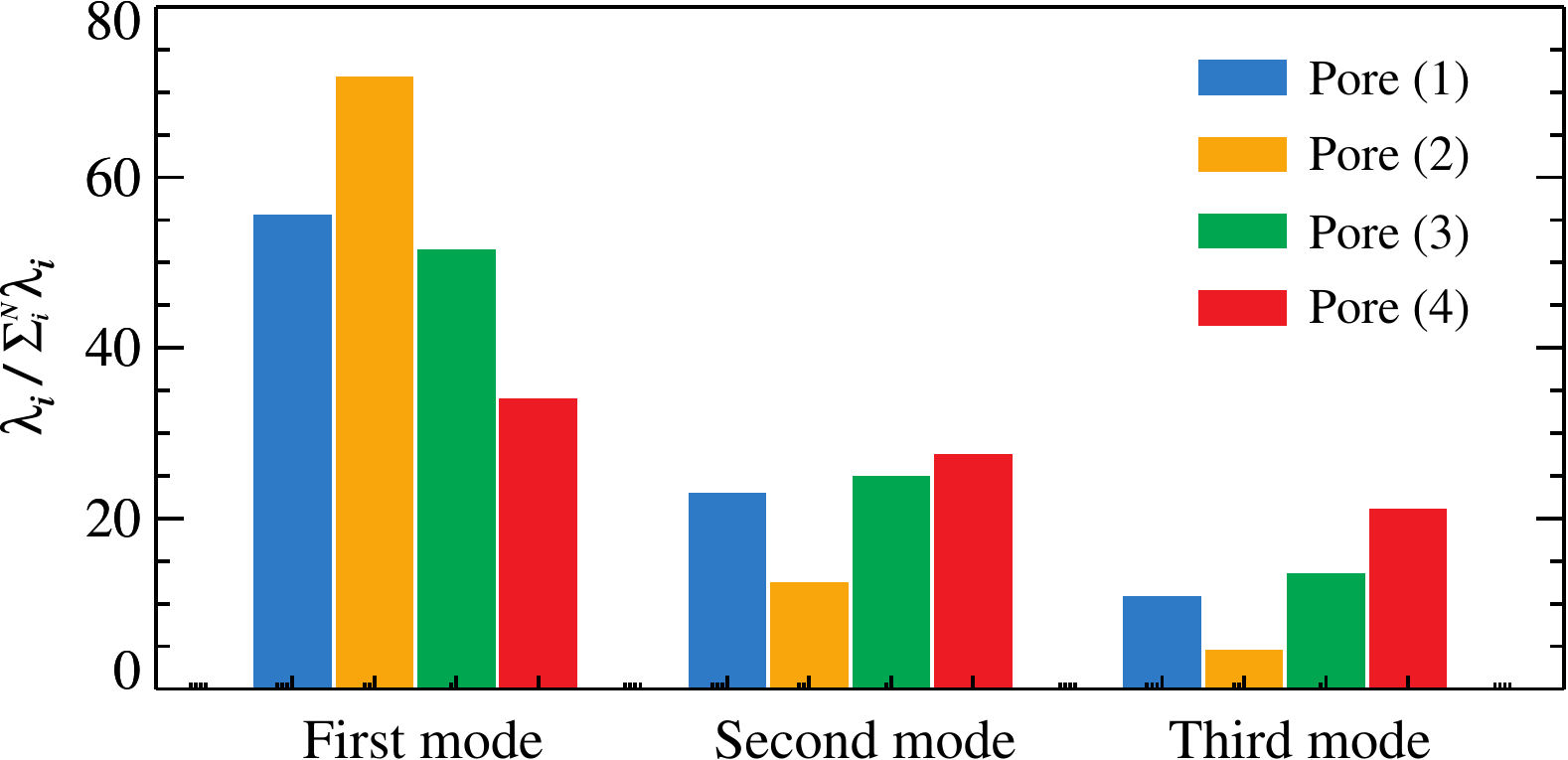}
         \caption{Percentage of eigenvalues, $\lambda_i$, normalised by the total sum of all eigenvalues, $\Sigma_{i=1}^{N} \lambda_i$, of the first three POD modes. The first and second modes respectively correspond to MHD sausage and kink modes for all pores. The third POD mode were identified as a MHD fluting mode for pore (1), and kink modes for the other three pores.}
         \label{fig:pod_modes_energies}
\end{figure}

Finally, the eigenvalues, $\lambda_i$, associated with each POD mode $i$, normalised by the total sum of all eigenvalues, $\Sigma_{i=1}^{N} \lambda_i$, based on which the POD modes were ranked, are presented in Figure \ref{fig:pod_modes_energies}. This normalisation provides a metric for assessing the individual contribution of each mode to the average variance or oscillation of the pore boundary. The oscillations observed in the shapes and areas of pores (1)-(3) are mostly due to sausage modes, respectively responsible for 56\%, 72\%, and 52\% of the pores dynamics.
These are followed by kink modes with collective contributions (from the second and third POD modes) on the order of 23\%, 17\%, 39\% of the total eigenvalue, respectively. Pore (1) has additionally a contribution from a fluting mode, equal to 11\%. For pore (4), however, the kink modes together dominate over the sausage mode (49\% of the total eigenvalue versus 34\%, respectively).

\subsection{Phase-lag analysis}
\label{sec:[phase-analysis]}

Since the sausage modes were found to dominate the concurrent waves identified in the majority of the magnetic pores under study (three out of four), we perform an additional investigation by means of a phase-lag analysis. Phase differences between perturbations in various pairs of observables may, in comparison with theoretical models, facilitate further characterisation of different sausage wave modes \citep[see, e.g.][for a recent similar analysis]{2021RSPTA.37900175N}.

Using an MHD approach, \citet{2013A&A...551A.137M} determined the phase relationships between various parameters, namely (flux-tube averaged) intensity, LOS velocity, and LOS magnetic-field perturbations, in a uniform straight magnetic cylinder under solar photospheric conditions, with a particular focus on identifying different sausage modes. Hence, the various phase differences (summarised in Table~1 of their article) could suggest the presence of slow propagating/standing or fast propagating/standing surface modes. We use that table to investigate such characteristics in the four pores of interest studied here. Additionally, phase differences between area and intensity may reveal the slow/fast nature of the sausage modes, with in-phase relationships indicating slow modes, and anti-phase specifying fast modes \citep{2013A&A...555A..75M}.

\begin{figure}[htp]
     \centering
         \includegraphics[width=0.49\textwidth]{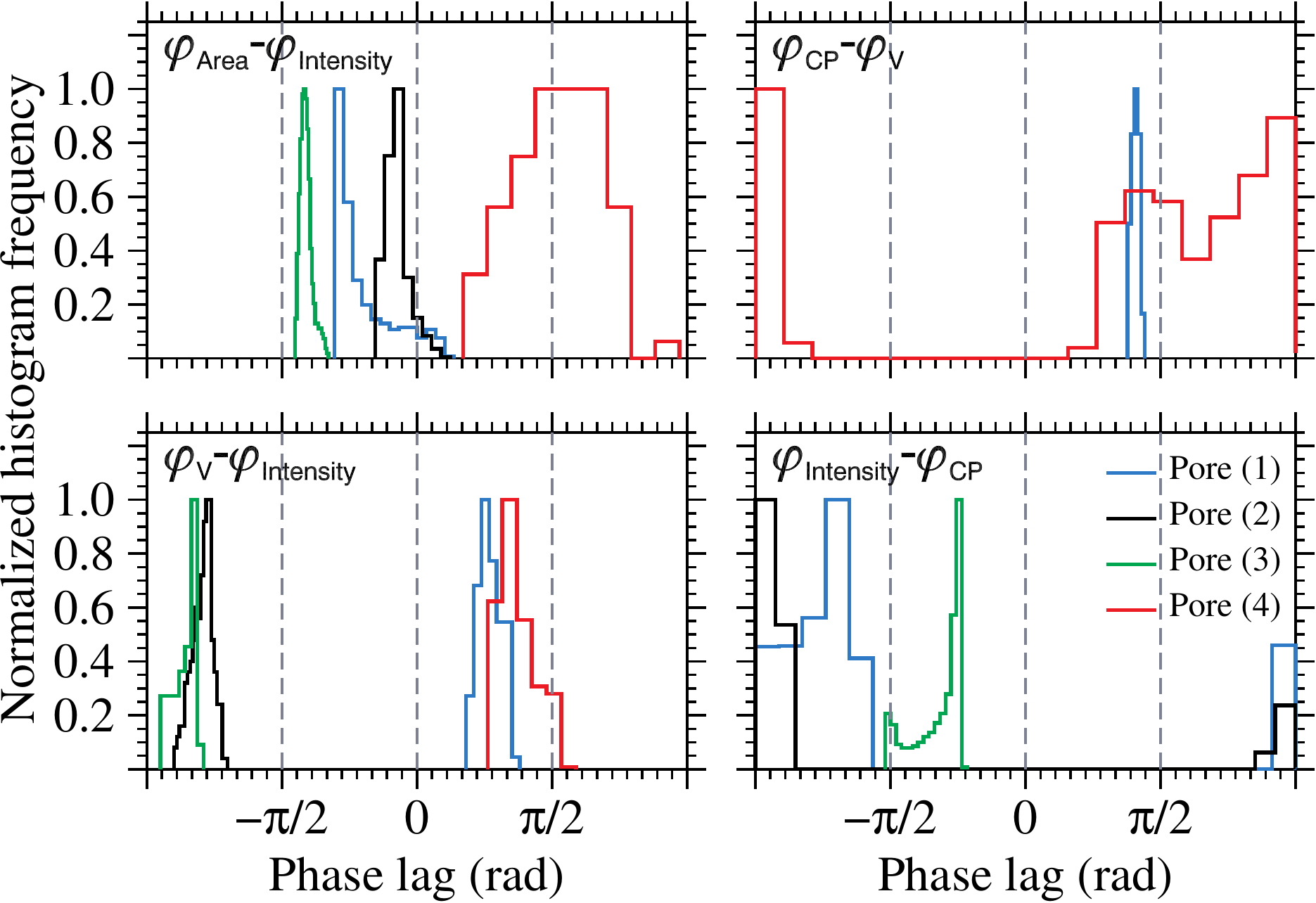}
         \caption{Distributions of phase lag for oscillations in four pairs of observables: Area--Intensity (top left), CP--LOS velocity (top right), LOS velocity--Intensity (bottom left), and Intensity--CP (bottom right), associated with the frequencies of the POD sausage modes. The histograms are normalised to their maximum values. The vertical lines mark  phase lags of $-\pi/2$, $0$, and $\pi/2$ radians.}
         \label{fig:phasediagram}
\end{figure}

We compute similar phase relationships as in \citep{2013A&A...555A..75M} and \citet{2013A&A...551A.137M}. These are obtained by calculating a wavelet coherence spectrum and phase differences between oscillations in the pairs of observables, that are Area--Intensity, CP--V, V--Intensity, and Intensity--CP for each pore (V represents the LOS velocity; positive V indicates red shift). From each wavelet coherence spectrum, distribution of the phase lags are obtained in regions with significant coherence (i.e. significant at 5\%; 95\% confidence level), being outside the wavelet's CoI. Figure~\ref{fig:phasediagram} illustrates distributions of phase lag, $\varphi$, for $\varphi_{area}$--$\varphi_{Intensity}$, $\varphi_{CP}$--$\varphi_{V}$, $\varphi_{V}$--$\varphi_{Intensity}$ , and $\varphi_{Intensity}$--$\varphi_{CP}$, retrieved from the four pores. For each pore, only phase relationships associated with the peak frequencies of the sausage modes obtained from the POD analyses (with a window of 1~mHz) entered the histograms. Positive phase lags indicate that the first parameter leads the second one (for details on how to determine such phase differences from wavelet coherence spectra, see \citet{2017ApJS..229...10J}). If no distribution is plotted for a particular pore in any of the phase relationships, that means no significant coherence was found between the corresponding fluctuations in the pairs of observables.

A comparison between these phase lags and those predicted in Table~1 of \citet{2013A&A...551A.137M} obtained in the case of an ideal flux tube is not straightforward and should be interpreted with great caution. One reason is the presence of various phase-lags (likely due to superposition of multiple wave modes in the same magnetic structure), so that a simple model may not account for such a complex situation as found in the real Sun. Additionally, the observed data may still contain noise and spurious signals contributing in such phase differences. 
These are difficult to distinguish from real signals. 
Furthermore, the observables are likely formed at slightly different geometric heights. Such height differences may additionally make the comparison of phase relationships between the observed parameters and those from the theoretical models (which were calculated at the same geometric heights) difficult. Thus, we primarily aim to present the phase-lags between the various observables and only attempt to tentatively make such comparisons for frequencies at which the sausage modes were identified by POD (i.e. those shown in Figure~\ref{fig:phasediagram}).

The best agreements with the theoretical models developed by \citet{2013A&A...551A.137M} were found for the $\varphi_{CP}$--$\varphi_{V}$ and $\varphi_{Intensity}$--$\varphi_{CP}$ phase relationships (from pores (4) and (3), respectively) implicating the identification of fast propagating surface modes. Additionally, the $\varphi_{Intensity}$--$\varphi_{CP}$ relationship (and to a lesser extend the $\varphi_{Area}$--$\varphi_{Intensity}$) for pore (2) suggested the presence of a slow sausage mode. However, since all distributions cannot show an agreement with those predicted by the models (see the previous paragraph for some possible reasons), such comparisons are not considered conclusive.

\section{Discussion and Conclusions}
\label{sec:conclusions}

We have inspected the concurrent presence of MHD wave modes in small magnetic pores captured by SO/PHI-HRT under seeing-free observing conditions. While we have studied perturbations in various observables, the focus has been on analysing oscillations of the area and shape of the pores.  The decomposition of these oscillations with the POD technique (for the first time to our knowledge) has revealed the existence of concurrent sausage and kink modes in all four pores under investigation. More interestingly, we have also identified a higher order eigenmode, a fluting mode, in one of these small-scale magnetic structures.

The POD analysis resulted in different distributions of eigenvalue for the four magnetic pores. While the dynamics in pores (1)--(3) was found to be dominated by the sausage modes, the kink modes had a larger contribution to the total eigenvalue in pore (4). Moreover, the fluting mode was identified only in pore (1). These differences could likely be due to the different local plasma (and magnetic) environments that the pores resided in. It is worth noting that the embedded plasma flows in different sections of the same active region have been shown to influence differently the kinematics of magnetic structures \citep{2017ApJS..229....8J}, so that they can affect the mode generation.  

We note that the MHD wave modes are identified/interpreted by the dominant characteristic of each POD mode (as a result of decomposition of the boundary oscillations), i.e. when their main motions are compared with cylindrical flux-tube models. However, the observed waveguides are irregularly shaped and may present multiple distortions while oscillating, as a result of, e.g inhomogeneous plasma inside and/or outside the magnetic structures or varying granulation buffeting from different angles. Thus, each POD mode could also present a flavour of another MHD wave mode due to, e.g. mode mixing, though not dominating. In fact, each POD mode could be a linear combination of MHD modes as the POD modal basis is estimated from the data themselves.

Although the characteristics of the local plasma/magnetic environment could be linked to the detected wave modes, we should also note that these identifications could perhaps be influenced by several other factors, such as the spatial and temporal resolutions, as well as the signal-to-noise ratio of the observations. In particular, while both the spatial and temporal resolutions can limit the detection of higher frequency fluctuations, the former may also create uncertainties in the pore's boundary detection. As such, the absence of other modes in our study does not necessarily imply that they are not present in the studied pores.

The most energetic mode in all four pores (i.e. that with the largest eigenvalue) was found to be the sausage mode with frequencies in the interval 1.5--3.0~mHz. Albeit these frequencies overlap, they are slightly smaller, on average, compared to those reported in the literature for photospheric magnetic pores, for instance, 2.2--33.3~mHz \citep{2011ApJ...729L..18M}, 2.4--5.5~mHz \citep{2015ApJ...806..132G}, and 2--12~mHz \citep{2018ApJ...857...28K}. \citet{2014A&A...563A..12D} reported frequencies of 1.2--3.7~mHz in area perturbations of a large pore, which have the largest overlap with those found here. However, we should note that frequencies of the area perturbations do not necessarily represent the frequencies of sausage-modes only, but could also include other wave modes. This also implies that by assuming all variations in area to be due to sausage modes, the sausage mode energy is overestimated, if the area changes are due to multiple wave modes.

The kink modes, as the second most energetic empirical mode (i.e. with the second largest eigenvalue) detected by POD was identified in a slightly smaller frequency range of 1.4--2.3~mHz (compared to that of sausage modes) in the four pores under study. Their frequencies are in the same order, but slightly lower, than the kink modes identified as the third empirical modes by POD (ranging between 1.8--2.7~mHz). The fluting mode was also found with a peak frequency at around 1.8~mHz. To the best of our knowledge, these are the first detections of kink and fluting modes in solar magnetic pores.
The similar frequencies across the identified MHD wave modes may suggest they are coupled to each other. 

The second and third POD spatial modes for pores (2)--(4) represented kink modes whose directions of fluctuations were perpendicular to each other. This may provide an intriguing development in relation to the search for the signatures of torsional Alfv\'en waves in the solar atmosphere. Observations of Alfv\'en waves have long been sought after due to their implications for atmospheric heating, however their incompressible nature makes detection of the velocity excursions around a magnetic flux tube boundary due to their torsional oscillations difficult \citep{2022SoPh..297..154C}. Recent studies have reported torsional velocity signatures in magnetic structures as evidence of Alfv\'en waves \citep[e.g.][]{2017NatSR...743147S, 2020A&A...633L...6K, 2021NatAs...5..691S}. However, the detection of orthogonal kink modes in this work implies that these modes in tandem may replicate the torsional behaviour of Alfv\'en waves \citep[cf. kink wave's rotational motions may appear similar to those expected from torsional Alfv\'en waves,][]{2014ApJ...788....9G}. Modelling of the nature of these modes with realistic drivers \citep[e.g.][]{2020A&A...639A..45K, 2021A&A...648A..77R} is vital to ascertain whether such kink mode interaction may influence the detection of Alfv\'enic motions in the solar atmosphere, particularly for the pores where the kink modes dominate over the sausage mode.

The large variations in pore area, up to $40\pm10$\%, may suggest a non-linear regime, with the most likely interpretation as a fast sausage mode according to the criteria developed by \citet{2013A&A...555A..75M}. Non-linear generation of fluting perturbations by kink modes have previously been predicted in both straight and twisted magnetic flux tubes \citep{2017SoPh..292..111R, 2018ApJ...853...35T, 2022SoPh..297..116R}. However, we should note that higher resolution observations would be needed to better clarify both the non-linearity and fast/surface nature of the sausage modes in such small-scale magnetic structures. 

To summarise, the concurrent sausage, kink, and fluting modes in photospheric small-scale magnetic structures have reliably been identified in the seeing-free data from SO/PHI-HRT. 
Understanding the behaviour and properties of these various MHD wave modes can provide us with valuable insights into the energy transfer mechanisms and the intricate dynamic processes occurring in the solar photosphere.
Application of POD on, for instance, CP and LOS velocity oscillations, in the inner structure of the pores could better quantify their specifications, that is the subject of a forthcoming article.
In addition, the propagation of the detected wave modes into the upper solar atmosphere would need further investigations using multi-height and multi-instrument diagnostics. 
Furthermore, we note that both POD and phase-lag analyses would greatly benefit from (a) a higher spatial- and temporal-resolution observations where the physical parameters within the magnetic structures were spatially and temporally resolved, and (b) the parameters, such as the magnetic field and LOS velocity, were inferred with a higher accuracy (i.e. with a higher spectral sampling/resolution). 
Hence, in future studies, we hope to conduct similar analyses on higher resolution observations \citep[from, e.g. the next flight of the {\sc Sunrise} balloon-borne solar observatory;][]{2010ApJ...723L.127S, 2017ApJS..229....2S} as well as on MHD simulations.
Additionally, longer seeing-free observations from SO/PHI-HRT in future observing campaigns (resulting in a higher frequency resolution) as well as future improvements on the HRT data-reduction routines (thus a lower noise level) are essential for identification of a larger number of MHD wave modes. More importantly, the spatial resolution of SO/PHI-HRT during the second (and subsequent) perihelion passes will be up to 60\% higher (minimum distance from the Sun $\le0.3$~au) compared to those analysed here (where the spacecraft was at about 0.5~au during this first perihelion passage).

\section*{Acknowledgements}
Solar Orbiter is a space mission of international collaboration between ESA and NASA, operated by ESA. We are grateful to the ESA SOC and MOC teams for their support. 
The German contribution to SO/PHI is funded by the BMWi through DLR and by MPG central funds. 
The Spanish contribution is funded by AEI/MCIN/10.13039/501100011033/ (RTI2018-096886-C5, PID2021-125325OB-C5, PCI2022-135009-2) and ERDF ``A way of making Europe''; ``Center of Excellence Severo Ochoa'' awards to IAA-CSIC (SEV-2017-0709, CEX2021-001131-S); and a Ram\'{o}n y Cajal fellowship awarded to DOS. 
The French contribution is funded by CNES. 
LACAS and PKB were supported by the Science and Technology Facilities Council (STFC, UK), grant ST/T00035X/1. LACAS also acknowledge support by STFC grant ST/X001008/1.
DBJ and SDTG acknowledge support from the UK Space Agency for a National Space Technology Programme (NSTP) Technology for Space Science award (SSc~009). DBJ and SDTG are also grateful to the UK Science and Technology Facilities Council (STFC) for additional funding via the grant awards ST/T00021X/1 and ST/X000923/1. DBJ also wishes to thank The Leverhulme Trust for grant RPG-2019-371. 
VF, GV and SSAS are grateful to the STFC for grant ST/V000977/1 and the Institute for Space-Earth Environmental Research (ISEE, International Joint Research Program, Nagoya University, Japan). VF and GV thank the Royal Society, International Exchanges Scheme, collaborations with Pontificia Universidad Catolica de Chile, Chile (IES/R1/170301). 
RG acknowledge the support by Funda\c{c}\~{a}o para a Ci\^{e}ncia e a Tecnologia (FCT) through the research grants UIDB/04434/2020 and UIDP/04434/2020. 
VF, GV, IB, SSAS thank the Royal Society, International Exchanges Scheme, collaborations with
Aeronautics Institute of Technology, Brazil, (IES/R1/191114), Monash University, Australia (IES/R3/213012), Instituto de Astrofisica de Canarias, Spain (IES/R2/212183),  Institute for Astronomy, Astrophysics, Space Applications and Remote Sensing, National Observatory of Athens, Greece (IES/R1/221095), and Indian Institute of Astrophysics, India (IES/R1/211123)
for the support provided. This research has also received financial support from the European Union's Horizon 2020 research and innovation program under grant agreement No. 824135 (SOLARNET). Finally, we wish to acknowledge scientific discussions with the Waves in the Lower Solar Atmosphere (WaLSA; \href{https://www.WaLSA.team}{https://www.WaLSA.team}) team, which has been supported by the Research Council of Norway (project no. 262622), The Royal Society \citep[award no. Hooke18b/SCTM;][]{2021RSPTA.37900169J}, and the International Space Science Institute (ISSI Team~502). 

\bibliographystyle{aa}
\bibliography{ref.bib}

\end{document}